\documentclass[%
preprint,
superscriptaddress,
amsmath,amssymb,
aps,
]{revtex4-1}

\usepackage{graphicx}
\usepackage{dcolumn}
\usepackage{bm}

\usepackage{graphicx}
\usepackage{bm}
\usepackage{amssymb}
\usepackage{dcolumn}
\usepackage[utf8]{inputenc}
\usepackage{subfigure}
\usepackage{threeparttable}
\usepackage{multirow}
\usepackage[usenames,dvipsnames]{color}
\usepackage{wasysym}
\usepackage{txfonts}
\usepackage{mathtools}

\begin{document}

\newcommand{\vn}[1]{{\bf{#1}}}
\newcommand{\vht}[1]{{\boldsymbol{#1}}}
\newcommand{\matn}[1]{{\bf{#1}}}
\newcommand{\matnht}[1]{{\boldsymbol{#1}}}
\newcommand{\bege}{\begin{equation}}
\newcommand{\ee}{\end{equation}}

\preprint{APS/123-QED}

\title[Sklenar \textit{et al}.]{Driving and detecting ferromagnetic resonance in insulators with the spin Hall effect}

\author{Joseph Sklenar}
\affiliation{Materials Science Division, Argonne National Laboratory, Argonne IL 60439, USA}
\affiliation{Department of Physics and Astronomy, Northwestern University, Evanston IL 60208, USA}

\author{Wei Zhang}
\affiliation{Materials Science Division, Argonne National Laboratory, Argonne IL 60439, USA
}%

\author{Matthias ~B.~Jungfleisch}
\affiliation{Materials Science Division, Argonne National Laboratory, Argonne IL 60439, USA}

\author{Wanjun~Jiang}
\affiliation{Materials Science Division, Argonne National Laboratory, Argonne IL 60439, USA}

\author{Houchen Chang}
\affiliation{Department of Physics, Colorado State University, Fort Collins CO 80523, USA}

\author{John E. Pearson}
\affiliation{Materials Science Division, Argonne National Laboratory, Argonne IL 60439, USA}

\author{Mingzhong Wu}
\affiliation{Department of Physics, Colorado State University, Fort Collins CO 80523, USA}

\author{John B. Ketterson}
\affiliation{Department of Physics and Astronomy, Northwestern University, Evanston IL 60208, USA}

\author{Axel Hoffmann}
\affiliation{Materials Science Division, Argonne National Laboratory, Argonne IL 60439, USA}

\date{\today}

\begin{abstract}
We demonstrate the generation and detection of spin-torque ferromagnetic resonance in Pt/Y$_\mathrm{3}$Fe$_\mathrm{5}$O$_\mathrm{12}$ (YIG) bilayers.  A unique attribute of this system is that the spin Hall effect lies at the heart of both the generation and detection processes and no charge current is passing through the insulating magnetic layer.  When the YIG undergoes resonance, a dc voltage is detected longitudinally along the Pt that can be described by two components.  One is the mixing of the spin Hall magnetoresistance with the microwave current.  The other results from spin pumping into the Pt being converted to a dc current through the inverse spin Hall effect.  The voltage is measured with applied magnetic field directions that range in-plane to nearly perpendicular.  We find that for magnetic fields that are mostly out-of-plane, an imaginary component of the spin mixing conductance is required to model our data.
\end{abstract}

\pacs{Valid PACS appear here}
\keywords{Suggested keywords}
\maketitle

Magnetic insulators such as Y$_\mathrm{3}$Fe$_\mathrm{5}$O$_\mathrm{12}$ (YIG) with extremely low magnetic damping serve as promising platforms for low power data transmission \cite{sspwu, wang_prb, houchen_ieee, pirro_apl}. In YIG/Pt bilayers the groundbreaking discovery of magnetization dynamics generated by spin orbit torques of Pt contacts \cite{kajiwara} opens up new opportunities for device concepts combining electronic, spintronic, and magnonic approaches. The spin orbit torques in heavy metals arise from the spin Hall effect (SHE) \cite{yakonov, hirsch_prl}, which converts a charge current, $\mathrm{\textbf{J}}_c$, to a spin current, $\mathrm{\textbf{J}}_s$, with a conversion efficiency dictated by a materials specific parameter, i.e., the spin Hall angle, ${\Theta}_{SH}$ \cite{hoffmann_ieee}.  The resultant spin current can drive spin-torque ferromagnetic resonance (ST-FMR) in bilayer thin films made from metallic ferromagnets and nonmagnetic metals \cite{liu_prl}. In such experiments, FMR is driven by the simultaneous Oersted field and oscillating transverse spin current (spin-torque) transformed by SHE from the alternating charge current. Electrical detection is made possible via the spin-torque diode effect \cite{sankey_prl}, i.e., the rectification of the time dependent bilayer resistance arising from the anisotropic magnetoresistance (AMR) of the ferromagnet \cite{juretschke_jap,dyakonov_prl}. However, such a detection scenario is not possible in magnetic insulators due to missing free electrons coupling to magnetic moments and, thus, the absence of AMR.

In this Letter, we show experimentally that the SHE of a paramagnetic metal can be used for both excitation and detection of ST-FMR for magnetic insulators. We demonstrate magnetization dynamics of a thin YIG layer induced by spin-torque from an adjacent Pt layer, as well as subsequent detection of a dc voltage via the spin-torque diode effect generated by the anisotropic spin Hall magnetoresistance (SMR) of the Pt \cite{nakayama_prl,chen_prb,althammer_prb,dyakonov_prl, hahn_prb}.  It bears mentioning that the anisotropic resistance of metal films on top of ferromagnetic insulators, and interface effects in general \cite{rojas_prl}, are a very active topic and other mechanisms independent of the SHE such as interface proximity effects \cite{miao_prl} and interfacial Rashba effects \cite{grigoryan_prb} are being explored as contributors.  In this work, SMR refers to the dependence of the electrical resistance of the metal on the magnetization direction of an adjacent magnetic insulator and is a result of a simultaneous operation of the SHE and its inverse (ISHE) as a nonequilibrium phenomenon.  Microscopically, this anisotropic behavior orginates from the dependence of the spin accumulations of conduction electrons at the YIG/Pt interface on the static YIG magnetization.  For example, if the static magnetization is aligned with the spin current's polarization at the interface there is a large backflow \cite{jiao_prl} spin current; on the other hand, if the magnetization is orthogonal to the polarization a spin current is absorbed at interface, and consequently the interfacial spin accumulation is reduced.  

Models of spin transport at the YIG/Pt interface that exclude proximity effects introduce the spin mixing conductance, $G^{\uparrow\downarrow}$, to describe both the magnitude and phase of the interface spin current \cite{tserkovnyak_prl}.  This concept has been probed in a comprehensive study \cite{weiler_prl} involving a suite experiments such as spin pumping \cite{mosendz_prb,bai_prl}, spin Seebeck detection \cite{kikkawa_prl}, and SMR measurements.  It has also been shown that the value of $G^{\uparrow\downarrow}$ for a YIG/Pt interface is heavily dependent on sample fabrication and processing \cite{benjamin_apl}.  In these works the spin mixing conductance is typically described as being purely real.  However, for YIG/Pt bilayers it has been theoretically suggested that a non-zero value of Im($G^{\uparrow\downarrow}$) should be considered \cite{chiba_jap}.  Furthermore, very recent experiments investigating an anamolous spin Hall effect in Pt have provided evidence for a non-zero Im($G^{\uparrow\downarrow}$) at the YIG/Pt interface \cite{meyer_arx}.   Here, we will present evidence that for ST-FMR experiments where the magnetic field is tipped out-of-plane (OOP) a non-zero Im($G^{\uparrow\downarrow}$) is required and evolves as a function of the OOP angle.

\begin{figure}
\includegraphics{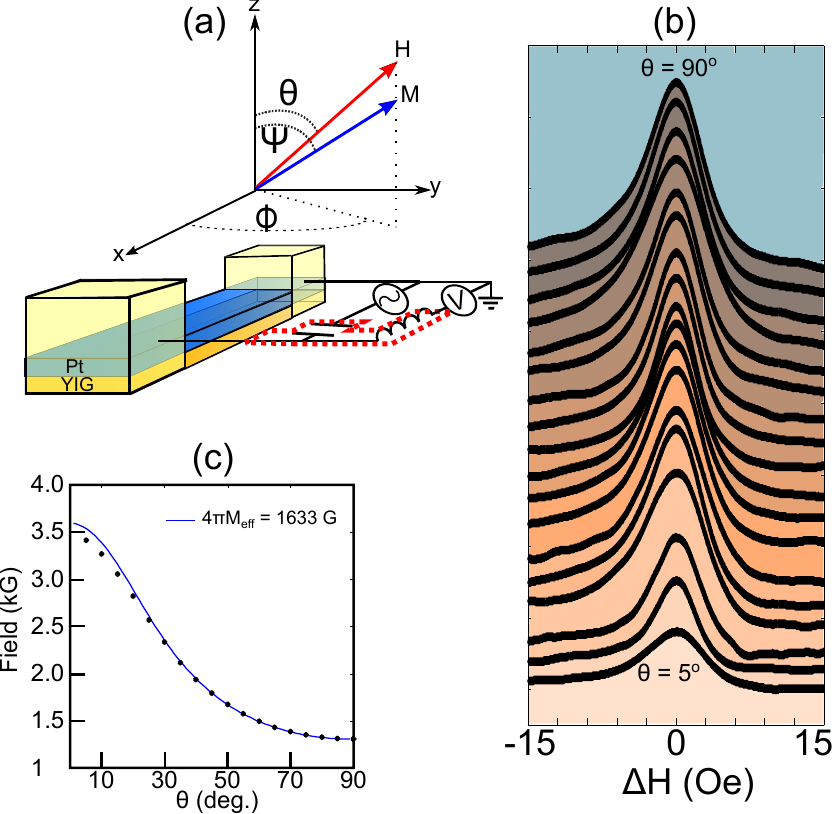}
\caption{A schematic of the bilayer and ST-FMR set-up is shown in (a).  In the diagram \textbf{H} indicates an experimentally applied field, and \textbf{M} indicates the magnetization vector.  $\theta$ describes the tipping of \textbf{H} from the z-axis (thickness direction) and $\psi$ describes the tipping of \textbf{M} in the same manner. $\phi$ is an in-plane angle between the x and y axis; in all our experiments $\phi$ = 45$^\circ$.  (b) ST-FMR traces measured over a range of $\theta$ that spans from 90$^\circ$ - 5$^\circ$ in 5$^\circ$ steps.  In order to show every resonance we plot each resonance centered on zero field.  (c) shows the $\theta$ dependence of the ST-FMR experiments fit to Eq. (4).  4$\pi$M$_{eff}$ is extracted from this data set to be 1633 G.}
\label{fig:Appar}
\end{figure}

We fabricated YIG(40 nm)/Pt(6 nm) bilayers by \textit{in-situ} magnetron sputtering on single crystal gadolinium gallium garnet (GGG, Gd$_\mathrm{3}$Ga$_\mathrm{5}$O$_\mathrm{12}$) substrates of 500 $\mu$m thickness with [111] orientation under high-purity argon atomsphere \cite{houchen_ieee,tao_jap}.  The bilayers were subsequently patterned into microstripes in the shape of 500 $\mu$m $\times$ 100 $\mu$m by photolithography and liquid nitrogen cooled ion milling to remove all the YIG/Pt materials except for the bar structure.  In a last fabrication step, square contact pads made of Ti/Au (3 nm / 120 nm) are patterned on top each end of the YIG/Pt stripe via photolithography and lift-off.  We configured our set-up into a ST-FMR scheme that is illustrated in Fig.~\ref{fig:Appar} (a).  A bias-tee is utilized to allow for simultaneous transmission of microwaves as well as dc voltage detection across the Pt.   We modulate the amplitude of the microwave current at 4 kHz so that the ST-FMR dc signal is detected via a lock-in amplifier to improve signal to noise.

The coordinate system that we will reference throughout this work is shown in Fig.~\ref{fig:Appar} (a).  The angle $\phi$ is in-plane and lies between the x and y axis.  For our experiments $\phi$ was always set to $45^\circ$.  The polar angle $\theta$ describes the applied magnetic field direction OOP, while the polar angle $\psi$ is the calculated OOP component of the magnetization.  Due to geometrical demagnetization fields $\psi > \theta$; for a given $\theta$ and applied magnetic field $\psi$ is determined from the following expression:
\begin{equation}
{
2\pi M_{eff}\sin2\psi\csc(\psi - \theta) - H_{ex} = 0,
}
\end{equation}
where $M_{eff}$ is the effective magnetization of the YIG and $H_{ex}$ is the externally applied magnetic field.

To induce ST-FMR in the YIG we passed a fixed 5.5 GHz signal through the Pt while sweeping $H_{ex}$ at a fixed $\phi$ and $\theta$.  The nominal microwave power level was set to be 10 dBm.  The dynamic response of the system is governed by a modifed LLG equation of motion \cite{chiba_jap}:
\begin{equation}
\frac{d\hat{\textbf{M}}}{dt} = -\mid\gamma\mid \hat{\textbf{M}} \times \textbf{H}_{eff} + \alpha_\circ \hat{\textbf{M}} \times \frac{d\hat{\textbf{M}}}{dt} + \frac{\mid\gamma\mid \hbar \textbf{J}_s}{2eM_sd_{F}},
\end{equation}
where $\textbf{H}_{eff}$ includes the Oersted field, $H_{ac}$, demagnetization fields, and the applied external dc field $H_{ex}$.  Additional quantities of importance are the intrinsic damping, $\alpha_{\circ}$ and the spin current at the interface,
\begin{equation}{
\textbf{J}_s = \frac{Re(G^{\uparrow\downarrow})}{e} \hat{\textbf{M}} \times (\hat{\textbf{M}} \times \mu_{s}) + \frac{Im(G^{\uparrow\downarrow})}{e} \hat{\textbf{M}} \times \mu_{s}
}
\end{equation}
that originates from the SHE in Pt.  Here $G^{\uparrow\downarrow}$ is the spin mixing conductance and $\mu_s$ is the spin accumulation at the YIG/Pt interface.  The oscillatory torque terms that drive the magnetization are the field from the microwave current in H$_{eff}$ and the spin torque term that includes $J_s$.  The angular range that $\theta$ covered over the course of our experiment was 5$^\circ$ - 90$^\circ$ in steps of 5$^\circ$.  Figure~\ref{fig:Appar} (b) plots every trace that was observed over the measureable angular range of $\theta$.  The OOP field dependence of the resonances shown in (b) is plotted in Fig.~\ref{fig:Appar} (c).  In order to extract the effective saturation magnetization of our YIG we fit (Fig.~\ref{fig:Appar} (c)) the out-of-plane angular dependence to the generalized Kittel equation that is given by:
\begin{equation}{
f = \frac{\mid\gamma\mid}{2\pi}\sqrt{H^2 + 4\pi M_{eff}(H(\sin\theta \sin\psi - 2\cos\theta \cos\psi) + 4\pi M_{eff}\cos^2\psi)},
}
\end{equation}
where $\gamma$ is the gyromagnetic ratio taken as 2.8 GHz/kOe.  The extracted effective magnetization is $4\pi M_{eff}$ = 1633 G.  We  note that this Kittel-like analysis does not account for magnetocrystalline anisotropy or exchange energy.  For comparison, in a separate work involving the study of spin waves in other thin YIG films we measured $4\pi M_{eff}$ = 1553 G \cite{benjamin_mmm}.

To explain our experimental observations, we employ a theory developed by Chiba \textit{et. al.} \cite{chiba_prap,chiba_jap}.  Qualitatively, this model desribes a dc voltage that develops longitudinally along the Pt film when a microwave charge current flowing through the Pt induces ferromagnetic resonance in the YIG.  There are two different contributions to the observed voltage: first, there is an analog to what is observed for Py/Pt bilayers where AMR of the Py mixes with the microwaves to generate a dc voltage at and near the FMR condition \cite{liu_prl}.  For YIG/Pt the magnetoresistance resides in the Pt and is the SMR \cite{nakayama_prl, chen_prb, althammer_prb}.  Additionally, spin pumping at the YIG/Pt interface can inject a spin current into the Pt that can be converted to a dc charge current via the ISHE.

The theoretical model \cite{chiba_prap,chiba_jap} predicts that the voltage generated by spin pumping has a purely symmetric lineshape about the resonance condition, and that the voltage induced by SMR also has a symmetric contribution.  Furthermore, the SMR contribution has an antisymmetric contribution to the lineshape as well.  This model \cite{chiba_prap} was recently expanded to include a non-zero imaginary part of G$^{\uparrow\downarrow}$, a phase shift parameter, $\delta$, between the charge current $J_c$ and $H_{ac}$, and an OOP applied dc Oersted field \cite{chiba_jap}.  $\delta$ should be considered to be a property of a given device and, for a fixed excitation frequency, should be constant.  The addition of the non-zero imaginary part of G$^{\uparrow\downarrow}$ along with the phase shift parameter $\delta$ allows for additional tunability in the net amplitude of both the antisymmetric as well as the symmetric contribution to the lineshape.

According to theory, the lineshapes of a ST-FMR experiment for a YIG/Pt bilayer have the following functional forms \cite{chiba_jap}:
\begin{multline}
V_{SMR} = [S_1F_S(H_{ex}) + A_1F_A(H_{ex})]\cos\phi \sin2\phi \sin\theta \\ - [S_2F_S(H_{ex}) + A_2F_A(H_{ex})]\sin^3\phi \cos\theta\sin2\theta \\ + A_3\sin\phi \sin2\phi \sin2\theta
\end{multline}

\begin{multline}
V_{SP} = S_3\cos\phi\sin2\phi\sin\theta + S_4\sin^3\phi\cos\theta\sin2\theta \\+S_5\sin\phi\sin2\phi\sin2\theta,
\end{multline}
where $V_{SMR}$ arises from SMR and $V_{SP}$ is from spin pumping.  $F_S(H_{ex})$ is the field dependent symmetric lineshape that is given by $\Delta^2/[(H_{ex} - H_{FMR})^2\cos^2(\theta - \psi) + \Delta^2]$.  $F_A(H_{ex})$ is an antisymmetric lineshape that is given by $F_S(H_{ex})\cos(\theta - \psi) (H_{ex} - H_{FMR})/\Delta$.  In these equations $\Delta$ is the linewidth of the lineshape and $H_{FMR}$ is the field under which FMR occurs, which can be obtained from inverting Eq. (3).  $S_1 - S_2$, and $A_1 - A_3$ are coefficients that rely on the mixing of the oscillatory SMR with the charge current, and all end up being proportional to $J_c^2$; the other relevant parameters such as $\Theta_{SH}$, G$^{\uparrow\downarrow}$, $\delta$, $M_{eff}$, $d_N$, and $d_F$, are imbedded within these coeffiencients \cite{chiba_jap}.  Two other parameters not yet mentioned are contained within these coefficients; they are the Pt resistivity $\rho$, and the spin diffusion length $\lambda$.  In our analysis we use $\lambda$ = 1.2 nm; this value was determined for Pt by spin pumping experiments in Py/Pt bilayers \cite{zhang_apl}. $S_3 - S_5$ are spin pumping coefficients that are similarily proportional to $J_c^2$ and depend on the same quantities listed above for the SMR terms.  Complete expressions for these coefficients can be found elsewhere \cite{chiba_jap}.

In our analysis there are three fitting parameters assumed to be independent of $\theta$: $\Theta_{SH}$, $J_c$, and $\delta$.  We did not directly assume that the magnitude or complex composition of $G^{\uparrow\downarrow}$ was independent of $\theta$.  Because we have previously measured the $\Theta_{SH}$ of Pt to be 0.09 we analyze our data with this value in mind \cite{zhang_apl}.  In other ST-FMR experiments the paramater $\delta$ has been assumed to be zero, therefore we will begin our discussion by following this example \cite{liu_prl, mellnik_nature}.  This leaves us with fixing the magnitude of $J_c$.  Because the magnitude of $G^{\uparrow\downarrow}$ is free we found various values of $J_c$ could be used with reasonable $G^{\uparrow\downarrow}$ counterparts.  In fact, these two parameters are strongly anti-correlated.  However, we found that a given $J_c$ \textit{does not} ensure that the magnitude of $G^{\uparrow\downarrow}$ remains constant over all $\theta$.  We typically see an increase in the magnitude of $G^{\uparrow\downarrow}$ as the field is tipped OOP.  The value of $J_c$ (9$\times$ 10$^8$ A$/$m$^2$) chosen here minimized the variation of $G^{\uparrow\downarrow}$ over $\theta$ which then stays within 10$\%$ of a mean value of 2.44 $\times$ 10$^{14}$ $\Omega^{-1}$m$^2$.

\begin{figure}
\includegraphics{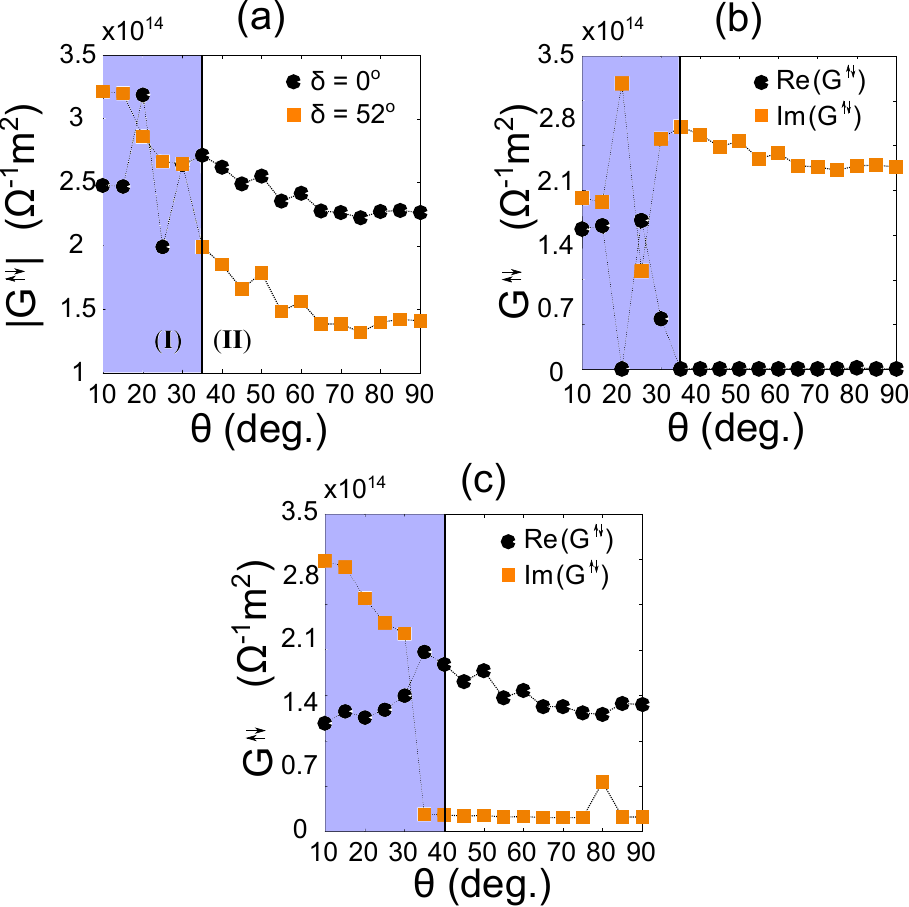}
\caption{The results of the $\theta$ dependence on both the real and imaginary components of the spin mixing conductance are shown above. In (a) $|G^{\uparrow\downarrow}|$ is plotted as a function of $\theta$ for two different assumed values of $\delta$.  The circles represent $\delta$ = 0$^\circ$ and the squares represent $\delta$ = 52$^\circ$.  In (b) the real and imaginary components of $G^{\uparrow\downarrow}$ are plotted as a function of $\theta$ for $\delta$ = 0$^\circ$.  In (c) the real and imaginary components are plotted for $\delta$ = 52$^\circ$.  } 
\label{fig:Angular1}
\end{figure} 

With $\Theta_{SH}$, $J_c$, and $\delta$ fixed we proceeded to investigate the magnitude and complex behavior of $G^{\uparrow\downarrow}$ as a function of $\theta$.  Fig. 2 (a) shows the $\theta$ dependence for our first set of assumptions as circles.  The complex behavior of $G^{\uparrow\downarrow}$ is plotted in Fig 2 (b) where the Re($G^{\uparrow\downarrow}$) is indicated as squares and the Im($G^{\uparrow\downarrow}$) is shown as circles.  Here, one sees that the composition of $G^{\uparrow\downarrow}$ is purely imaginary from $\theta$ = 35$^\circ$ - 90$^\circ$.  This  region is indicated as \textbf{II} in the plot.  For small values of $\theta$ ( $<$ 35$^\circ$) the composition begins to flucuate.  This region is indicated with a \textbf{I} and is shaded blue in Fig. 2.  As seen in Fig. 2 (b), for the smallest values of $\theta$, $G^{\uparrow\downarrow}$ settles on having real and imaginary components with similar magnitude.

Previously reported experiments, where the applied magnetic field is in-plane, report that $G^{\uparrow\downarrow}$ is mainly real, which is not consistent with our analysis.  A possible explanation may involve the parameter $\delta$.  In fact, $\delta$ has been used in a similar ST-FMR experiment where the in-plane field configuration and a near out-of-plane measurement was performed while $G^{\uparrow\downarrow}$ was assumed to be real \cite{schreier_arx}.  If we allow $\delta$ to vary we find that for a value of $\delta$ = 52$^\circ$ we had a local maximum in the ratio of Re($G^{\uparrow\downarrow}$)/$|G^{\uparrow\downarrow}|$, at $\theta$ = 90$^\circ$, as a function of $\delta$.  Therefore, we believe that a large phase shift between the microwave current and the microwave field exists making the analysis with a non-zero $\delta$ more appropriate.  With this new value of $\delta$, and with the same value of $J_c$ and $\Theta_{SH}$ as before, we performed again the $\theta$ dependent analysis.  The dependence that $G^{\uparrow\downarrow}$ has on $\theta$ with this non-zero $\delta$ is shown in fig. 2 (b) plotted as squares.  Fig. 2 (c) shows the complex composition of $G^{\uparrow\downarrow}$ for this non-zero $\delta$.  In contrast to before, for region \textbf{II}, $G^{\uparrow\downarrow}$ is mostly real with little flucuation in the angular range $\theta$ = 35$^\circ$ - 90$^\circ$.  However this behavior does not persist; we again we see that in region \textbf{I}, where the field approaches a OOP configuration, both the real and imaginary part of $G^{\uparrow\downarrow}$ become appreciably non-zero.

One conclusion from the above discussion is that the parameter space used in fitting ST-FMR lineshapes in a YIG-Pt bilayer is not well enough constrained.  To illustrate this point we show the model's flexibility in Fig. 3.  Here, we have plotted the model predictions directly on top of the data for both the zero and non-zero $\delta$ analysis and we have also chosen representative traces from both region \textbf{I} and region \textbf{II}.  What does emerge is that \textit{independent of the assumptions} used, for $\theta$ $<$ 35$^\circ$ both a real and imaginary component of $G^{\uparrow\downarrow}$ are needed to fit that data.  Before summarizing we note that we analyzed our data under different assumed values of $\Theta_{SH}$ (not shown).  Smaller assumed values of $\Theta_{SH}$ require smaller values of $\delta$ to make $G^{\uparrow\downarrow}$ mainly real at $\theta$ = 90$^\circ$. Near $\Theta_{SH}$ = 0.06 no $\delta$ is required.  Regardless, we see the same flucuating behavior of the complex composition of $G^{\uparrow\downarrow}$ for small values of $\theta$.  

\begin{figure}
\includegraphics[scale = .9]{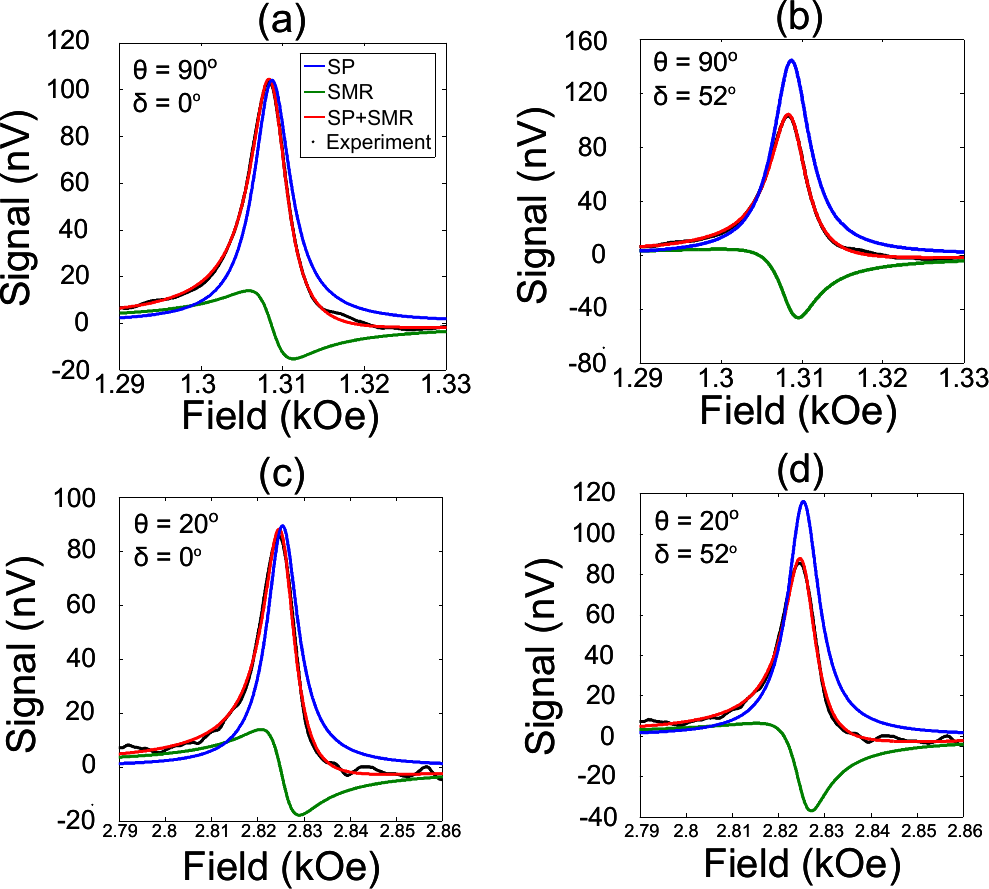}
\caption{Representative fits of the ST-FMR data for both zero and non-zero values of $\delta$.  Additionally, we show fits to the data for two different angles, $\theta$ = 90$^\circ$ and $\theta$ = 20$^\circ$.  These two angles each represent data acquired from regions \textbf{I} and \textbf{II} in fig. 2.  The black data points are densely packed together.  The total theoretical fit is plotted in red, while the two contributions to the total, spin pumping and SMR, are plotted in blue and green respectively. }
\label{fig:Angular2}
\end{figure} 
 
The ST-FMR paradigm has been studied with great intensity for spin Hall metal/ferromagnetic bilayers where the ferromagnet is a conductor.  The present work shows that it can be successfully extended to insulating FM materials.  Furthermore, it is clear that in addition to an Oersted microwave field torque from the Pt strip line, an additional spin torque from spin accumulation at the Pt/YIG drives the dynamics as well.  This particular conclusion is bolstered by a good agreement with theory that includes such spin torques.  A very interesting property of bilayers with ferromagnetic insulators such as YIG is that the longitudinal voltage generated along the Pt when ST-FMR is taking place is created by effects that all trace their origin back to the SHE.  These detection mechanisms set this work apart from metallic ferromagnets where mixing of the microwave current with the AMR of the ferromagnet itself leads to a measurable voltage.  In this work we have also have realized a recently proposed model \cite{chiba_jap} that describes ST-FMR voltages in YIG/Pt bilayers.  We highlight that in order to adequately model our data over the full angular range, the value of Im($G^{\uparrow\downarrow}$) was found to be an appreciable quantity for applied magnetic fields where the magnetization is sizably tipped OOP.

We acknowledge Stephen Wu for assistance with ion-milling used for sample prepartation.  The work at Argonne was supported by the U.S. Department of Energy, Office of Science, Materials Science and Engineering Division. Lithography was carried out at the Center for Nanoscale Materials, which is supported by DOE, Office of Science, Basic Energy Science under Contract No. DE-AC02-06CH11357.  Work at Northwestern utilized facilities maintained by the NSF supported Northwestern Materials Research Center under contract number DMR-1121262.  The work at Colorado State University was supported by the U. S. Army Research Office (W911NF-14-1-0501), the U. S. National Science Foundation (ECCS-1231598), C-SPIN (one of the SRC STARnet Centers sponsored by MARCO and DARPA), and the U. S. Department of Energy (DE-SC0012670).

\end{document}